\documentclass[12pt]{iopart}
\usepackage{iopams}
\usepackage[final,dvips]{graphicx}

\begin{document}

\title[Fluctuations in two-band superconductors]
{Fluctuations in two-band superconductors in a strong magnetic field}
\author{A. E. Koshelev$^1$ and A. A. Varlamov$^{1,2}$}

\address{$^1$ Materials Science Division, Argonne National Laboratory, 9700 S. Cass Ave.,
Argonne, Illinois 60439, USA}

\address{$^2$ CNR-SPIN, viale del Politecnico, 1 Roma 00133, Italia}

\ead{koshelev@anl.gov}

\begin{abstract}
We consider the behaviour of the fluctuating specific heat and conductivity in the
vicinity of the upper critical field line for a two-band superconductor.
Multiple-band effects are pronounced when the bands have very different coherence
lengths. The transition to superconductive state is mainly determined by the
properties of the rigid condensate of the ``strong'' band, while the ``weak'' band
with a large coherence length  of the Cooper pairs causes the nonlocality in
fluctuation behaviour and break down of the simple Ginzburg-Landau picture. As
expected, the multiple-band electronic structure does not change the functional
forms of dominating divergencies of the fluctuating corrections when the magnetic
field approaches the upper critical field. The temperature dependence of the
coefficients, however, is modified.  The large in-plane coherence length sets the field
scale at which the upper critical field has upward curvature.  The amplitude of
fluctuations and fluctuation width enhances at this field scale due to reduction of
the effective z-axis coherence length. We also observe that the apparent transport
transition displaces to lower temperatures with respect to the thermodynamic
transition. Even though this effect exists already in a single-band case at
sufficiently high fields, it may be strongly enhanced in multiband materials.
\end{abstract}
\maketitle

%Uncomment for PACS numbers title message
%\pacs{00.00, 20.00, 42.10}
%Keywords required only for MST, PB, PMB, PM, JOA, JOB?
%\vspace{2pc}
%\noindent{\it Keywords}: Article preparation, IOP journals
%Uncomment for Submitted to journal title message
%\submitto{\JPA}
%Comment out if separate title page not required

\section{Introduction}

Strong fluctuations is an important general feature of superconductors with high
transition temperature, small coherence length and high anisotropy. Thermal
fluctuations in superconductors is a mature field which has been developing for almost
fifty years. Quantitative analysis of fluctuations remains one of the best ways to
access the microscopic parameters of superconductors \cite{Tinkham,LVBook09}. The
recent microscopic theoretical studies of transport fluctuation phenomena for
arbitrary temperatures and magnetic fields
\cite{GL01,Lop0507,SSVG09,Fink09,GVV11a,TikhPRB12,TarasPRB13} allow us to analyze the
experimental findings for conventional and cuprate superconductors
\cite{Ong,Behnia06,Taill12,BatGl} in their phase diagrams along the upper-critical
field line $H_{c2}\left( T\right)$, in its periphery, and close to the quantum phase
transition at the point $H_{c2}(0)$.  Another important recent advance in the field
is development of quantitative theory accounting for strong fluctuations which
allows for accurate calculation of thermodynamic quantities in the vicinity of the
transition line, see Review \cite{RosensteinRMP10}.

Since the discovery of superconductivity in MgB$_{2}$ (see Review \cite{MgB2}) the
properties of multiband superconductors returned to the spotlight of attention after
half a century of oblivion \cite{2bandold}. Further discovery of multiband
high-temperature superconductivity in the iron pnictides and chalcogenides
%\cite{pnic}
gave an even stronger boost to this field, see recent experimental
\cite{PnictExpReviews} and theoretical \cite{PnictTheorReviews} reviews.

Superconducting properties of the magnesium diboride are strongly influenced by
multiband effects. Among the anomalies found in MgB$_{2}$ was the unusually narrow
temperature range of applicability of the standard Ginzburg-Landau (GL) theory
\cite{GLmgb2}. The Cooper pairs of different kinds, formed by the carriers of the
$\pi$ band and by the carriers of the $\sigma$ band respectively, behave themselves
as the unique condensate only very close to $T_{c}$. Due to the large difference in
the c-axis coherence lengths of $\sigma$ and $\pi$ bands, the condensates of
different kinds are already split at temperatures parametrically close to $T_{c}$:
$|T-T_{c}|/T_{c}\gtrsim\xi_{\sigma z}^{2} /\xi_{\pi z}^{2}+S_{\pi\sigma}\ll1$
\footnote{For parameters of MgB$_{2}$, $\xi_{\sigma z}^{2} /\xi_{\pi
z}^{2}+S_{\pi\sigma}\approx 0.02-0.05$.} (here $S_{\pi\sigma}\ll1$ is the relative
interband interaction constant).  Evidently, this particularity has to manifest
itself in fluctuations. The expected dominating effect is renormalization of the
effective c-axis coherence length which enters all fluctuation properties. While
near $T_c$ this length is given by the band-averaged value, above the crossover
temperature it is mostly determined by the more anisotropic $\sigma$ band and
therefore becomes much smaller.

Corresponding theory generalizing the microscopic theory of fluctuations on a
two-band superconductor and deriving the related nonlocal GL functional has been
developed in \cite{KVV05}. It was strongly focused on the applications to the
magnesium diboride in which the main differences between the bands are the strength
of intraband coupling constants and the values of the c-axis coherence length. In
result, the very early manifestation of the short wavelength fluctuations in the
$\pi$ band (where superconducting interaction is weaker) were predicted. These
predictions of the theory were actually never confirmed experimentally, and, in
general, an experimental situation  concerning fluctuations in MgB$_2$ is somewhat
uncertain. Early papers on the fluctuation conductivity \cite{SidorenkoJETP02},
magnetization \cite{LascialfariPRB02,MosqueiraPRB02}, and specific heat
\cite{ParkPRB02} described the experimental data using conventional single-band
theory. However, for analysis of the fluctuating magnetization very close to the
transition temperature $T_c$ in Ref.\ \cite{LascialfariPRB02}, an additional
artificial assumption on granular structure of the material was needed to describe
the data. In the follow up paper \cite{RomanoPRL05} the double-peak structure of the
fluctuation magnetization slightly above $T_c$ has been attributed to the two field
scales of the $\sigma$ and $\pi$ bands. To our knowledge, this is the only observed
fluctuation behaviour attributed to the two-band effects. The fluctuation
conductivity $\delta \sigma$ analyzed in Ref.\ \cite{putti} did not follow
theoretical predictions \cite{KVV05}. Instead, it had the typical two-dimensional
behaviour, $\delta \sigma \propto T_c/(T-T_c)$, in contradiction with the
three-dimensional electronic structure of the material.

The difficulties of detection of fluctuation multiband effects predicted by the
theory are probably related to the small amplitude of fluctuations in MgB$_{2}$. The
Ginzburg number in the clean limit is very small for this material, $Gi_{(3)}\approx
1.5\times10^{-6}\ $ \cite{KVV05}. Therefore it is challenging to fabricate uniform
samples with intrinsic transition width limited by fluctuations. Note that no
published fluctuation measurement has been done on a single crystal. Studies of the
fluctuation magnetization \cite{LascialfariPRB02,MosqueiraPRB02,RomanoPRL05} and
specific heat \cite{ParkPRB02} have been performed on misoriented powders or
polycrystalline samples and the fluctuation conductivity have been studied on thin
films \cite{SidorenkoJETP02,putti}.

In contrast to magnesium diboride, the iron pnictides are multiband semimetals and,
as a consequence, are characterized by quite strong fluctuations. Depending on
compound, the estimates for the Ginzburg number range from $3\times10^{-5}$ to
$5\times10^{-3}$ \footnote{These estimates are obtained for two iron-pnictide
compounds using the definition (\ref{GiDef}): BaFe$_{2}$(As$_{0.7}$P$_{0.3})_{2}$
($T_{c}=30$K, $\lambda_{xy}=274$nm, and $\xi_{z}=0.8$nm gives $Gi_{(3)}
\approx3\times10^{-5}$) and SmFeAsO$_{0.85}$F$_{0.15}$ ($T_{c}=50$K,
$\lambda_{xy}=136$nm, and $\xi_{z}=0.17$nm gives $Gi_{(3)}\approx
5\times10^{-3}$).}.
%\cite{pnictideGis}.
It is likely that behaviour of superconducting fluctuations in the iron-based
superconductors at sufficiently low temperatures and high magnetic fields is
influenced by the multiple-band effects. These effects are mostly pronounces when
there is a large difference between the gaps and coherence lengths in different
bands. Unfortunately, the partial coherence lengths for different bands are not
known in present. Behaviour of fluctuations has been analyzed for several iron-based
superconductors using conventional theory \cite{IronPnictFluct} and no obvious
multiband effects have been
reported so far. %
On the other hand, noticeable upward curvature in the temperature dependence of the
upper critical field observed in some compounds \cite{Hc2Tdep} has been attributed
to multiband effects, see also Reviews \cite{GurevichRPP11}. Also, it has been
observed that in FeSe$_{0.5}$Te$_{0.5}$ \cite{SerafinPRB10} and
BaFe$_{2}$(As$_{0.7}$P$_{0.3})_{2}$ \cite{WelpMarcenat} in strong magnetic field the
apparent transport transition is displaced to the lower temperatures relative to the
thermodynamic transition measured by the specific heat. It is plausible that this
displacement is also caused by multiband effects.
%On the other hand, there are preliminary data suggesting that multiband effects are
%noticeable in the fluctuation properties of the P-doped 122 compounds in strong
%magnetic field \cite{WelpMarcenat}. Similar behaviour was also observed in
%FeSe$_{0.5}$Te$_{0.5}$ \cite{SerafinPRB10}.

Usually the effect of a magnetic field on fluctuations becomes essential when the
magnetic length $\ell_{H}$ reaches the value of the fluctuation Cooper pair size.
Since the coherence lengths of different bands together with the gaps in a
multiple-band superconductor can differ strongly, one can expect that the
short-wavelength fluctuation modes in them will be excited at very different fields,
like it was found in the temperature dependencies of the paraconductivity and
fluctuation heat capacity for MgB$_{2}$ \cite{KVV05}.

In this article we consider thermal fluctuations in a two-band superconductor near
the upper critical field $H_{c2}(T)$ for arbitrary relation between the coherence
lengths and general structure of the coupling-constant matrix. Remaining in the
region of relatively strong magnetic fields along the line $H_{c2}(T)$, suitable for
analysis of experimental data, we will derive the general formulas for the
fluctuation corrections to the specific heat and conductivity.

\section{Fluctuations near the upper critical field for a single-band
superconductor}

Manifestations of fluctuations in the phase diagram of a superconductor are very rich
and diverse  \cite{Tinkham,LVBook09,SSVG09,GVV11a}. The line $H_{c2}\left( T\right)
$ can be approached in its initial part where $H_{c2}\left( T\right)  \ll
H_{c2}\left( 0\right)  $ within the frameworks of the GL theory. In this section we
summarize well-known results for the fluctuation specific heat and conductivity near
the upper critical field line. In the immediate vicinity of the transition
temperature these results are also valid for multiband superconductors. Since the
main modification of the theory for the two-band case includes a proper treatment of
fluctuations with the length scales below one of the microscopic coherence length,
we also briefly overview known manifestations of such short-scale fluctuations in
conventional single-band superconductors.

\subsection{Specific heat}

Fluctuation correction to the specific heat of the superconductor with the axial
symmetry of the spectrum placed in the magnetic field $H$ directed along the axis of
symmetry can be easily calculated in the frameworks of the GL scheme
\cite{LVBook09}. One can write the fluctuation contribution to the free energy in
the Landau representation as
%\begin{equation}
%F_{\mathrm{GL}}(\epsilon,H)\!=\!-{\frac{H{V}}{\Phi_{0}}T}\sum_{n=0}
%\int\limits_{-\infty}^{\infty}\frac{dq_{z}}{2\pi}\ln\frac{{\ \pi T}}{\alpha
%T_{\mathrm{c}}\epsilon\!+\!\omega_{\mathrm{c}}\left(  n\!+\!\frac{1}
%{2}\right)  \!+\!q_{z}^{2}/2m_{z}}. \label{ffe2mf}
%\end{equation}
\begin{equation}
F_{\mathrm{GL}}(\epsilon,H)\!=\!-{\frac{H V}{\Phi_{0}}T}\sum_{n=0}
\int
%\limits_{-\infty}^{\infty}
\frac{dq_{z}}{2\pi}\ln\frac{{A}}{\epsilon
\!+\!h\left(  2n\!+\!1\right)  \!+\!\xi_{z}^{2}q_{z}^{2}}.
\label{ffe2mf}
\end{equation}
Here ${V}$ is the sample volume, $\Phi_{0}=\pi\hbar/e$ is the magnetic flux quantum,
$\epsilon=\ln T/T_{c}$, $h=H/\tilde{H}_{c2}(0)$ is the reduced magnetic field,
$\tilde {H}_{c2}(0)=\Phi_{0}/\left(  2\pi\xi_{xy}^{2}\right) $ is the linear
extrapolation of the second critical field at zero temperature, $\xi_{xy}$ and
$\xi_{z}$ are the transversal and longitudinal coherence lengths.

%AEK16JUN: factor ${HS}/\Phi_{0}$ vanished
%, and the degeneracy of Landau levels results in appearance of the
%number of particle states ${HS}/\Phi_{0}$ with the definite quantum numbers
%$n$ and $k_{z}$.
The sum in equation (\ref{ffe2mf}) is evidently divergent and in order to
regularize it one should introduce a formal cut-off parameter, the number of
the last Landau level, at which the summation is interrupted (we address the
curious reader to \cite{LVBook09}). Since we are interested here in the
fluctuation heat capacity, i.e., the second derivative of the free energy (\ref{ffe2mf}),
the problem of regularization does not arise and one can write
\begin{eqnarray}
\delta C(\epsilon,H)  &  \approx-\frac{1}{VT}\frac{\partial^{2}}
{\partial\epsilon^{2}}F_{\mathrm{GL}}(\epsilon,H)\nonumber\\
&  ={\frac{h}{8\pi\xi_{xy}^{2}\xi_{z}}}\sum_{n=0}^{\infty}\frac{{\ 1}}{\left[
\epsilon+h\left(  2n+1\right)  \right]  ^{3/2}}, \label{ch}
\end{eqnarray}
where $h_{c2}(T)\equiv H_{c2}\left(  T\right)
/\tilde{H}_{c2}(0)=-\epsilon\approx(T_{c}-T)/T_{c}$.

Along the line $H_{c2}\left(  T\right)  $ the most singular contribution in the sum
(\ref{ch}) arises from the lowest Landau level $n=0$,
\begin{equation}
\delta C=\frac{1}{8\pi\xi_{xy}^{2}\xi_{z}}\frac{h_{c2}}{\left(  h-h_{c2}
\right)  ^{3/2}}. \label{C3D}
\end{equation}
Comparing it to the mean-field value of the heat capacity jump $\Delta C=8\pi^{2}\nu
T_{c}/\left(  7\zeta(3)\right)  $, one find that
\begin{equation}
\frac{\delta C}{\Delta C}=\sqrt{Gi_{(3)}}\frac{h_{c2}}{\left(  h-h_{c2}
\right)  ^{3/2}} \label{CDC}
\end{equation}
with the 3D Ginzburg number
\begin{equation}
Gi_{(3)}=\left[  \frac{7\zeta(3)}{64\pi^{3}}\frac{1}{\xi_{xy}^{2}\xi_{z}\nu
T_{c}}\right]  ^{2}=\frac{16\pi^{4}\lambda_{xy}^{4}T_{c}^{2}}{\xi_{z}^{2}
\Phi_{0}^{4}}, \label{GiDef}
\end{equation}
where $\lambda_{xy}$ is the in-plane London penetration depth, $\nu$ is the
normal-state density of states per spin, and $\zeta(3)\approx 1.202$. The results
(\ref{C3D}) and (\ref{CDC}) are valid for magnetic fields exceeding the typical
value $H_{Gi}\equiv\tilde{H}_{c2}(0)Gi_{(3)}$. As follows from (\ref{CDC}), the
fluctuation width of the superconducting transition in magnetic field grows as
$Gi_{(3)}(H)=Gi_{(3)}^{1/3}h^{2/3}$, see figure \ref{Fig-FluctWidth}(a).

\subsection{Paraconductivity}
\label{Sec-SigSnglBand}

Time-dependent Ginzburg-Landau theory allows to write near the initial part of the
line $H_{c2}\left(  T\right)  $ also the expression for paraconductivity, see, e.g.,
\cite{LVBook09},
\begin{equation}
\sigma_{xx}^{AL}=\frac{e^{2}}{8\xi_{z}}\frac{1}{\sqrt{h-h_{c2}}}.\label{sxx}
\end{equation}
Being normalized on the Drude conductivity of the normal phase, $\sigma
_{n}=2e^{2}\nu \mathcal{D}$, where $\mathcal{D}$ is the diffusion constant, this
result can be represented as
\begin{equation}
\frac{\sigma_{xx}^{AL}}{\sigma_{n}}=\frac{\pi^{4}}{14\zeta(3)}\sqrt{Gi_{(3)}
}\frac{1}{\sqrt{h-h_{c2}}}.\label{SLD}
\end{equation}
Comparing Eqs. (\ref{CDC}) and (\ref{SLD}), we see that the apparent widths of
transition are different for thermodynamics and transport. We can define the field
$h^{\ast}$ at which the fluctuation correction to heat capacity becomes of the order
of the jump, corresponding to the boundary of the fluctuation region,
$h^{\ast}-h_{c2}=Gi_{(3)}^{1/3}|\epsilon|^{2/3}$. At this field we have
\begin{equation}
\frac{\sigma_{xx}^{AL}}{\sigma_{n}}(h^{\ast})=\frac{\pi^{4}}{14\zeta
(3)}\left(  \frac{Gi_{(3)}}{|\epsilon|}\right)  ^{1/3}
\label{sigRatSnglBnd}
\end{equation}
with $\pi^{4}/[14\zeta(3)]\approx5.8$. We see that this ratio remains small at
$h=h^\ast$ for temperatures $|\epsilon|> \left(  \frac{\pi^{4}}{14\zeta(3)}\right)
^{3}Gi_{(3)} \!\sim \! 194\ Gi_{(3)}$ corresponding to fields $H> 194 H_{Gi}$. This
result implies that at sufficiently high fields the relative fluctuation correction
to conductivity remains small even when fluctuations of the order parameter become
strong and non-Gaussian.
%Recalling that for pnictides the Ginzburg-Levanyuk
%number is estimated as $3\times10^{-5}$ to $5\times10^{-3}$, one can notice, that
%the superconducting transition looking on the temperature dependence measured in
%strong enough magnetic field may look much more narrow than that one observed by
%resistive measurements.

\subsection{Short-scale fluctuations in conventional superconductors}

Historically, the first manifestation of the short wave-length fluctuations in
conventional superconductors has been revealed in the field dependence of the
fluctuation magnetization above $T_c$ about forty year ago \cite{Tinkham}. The GL
theory for a three-dimensional superconductor \cite{ASchmid,SchH,VS,Pr} predicts
that the fluctuation magnetization grows linearly with the magnetic field for $H
<\tilde{H}_{c2}(0)\epsilon$ and crosses over to the $\sqrt{H}$ dependence for higher
fields. Such a limitless growth of the fluctuation magnetization with field is
somewhat counter-intuitive, because one would expect that the magnetic field should
eventually suppress superconducting fluctuations. Indeed, the experimental
fluctuation magnetization  shows \emph{nonmonotonic} field dependence and at high
fields its values occur to be significantly lower than the GL prediction
\cite{Gollub,SotoPRB04}. This apparent contradiction has been resolved within the
microscopic theory which properly accounts for the short-wave-length and dynamical
fluctuations \cite{LeeP71,KAE72,MakiT}. It occurred that in the clean case the GL
theory breaks down very early, at magnetic fields of the order of $0.05 H_{c2}(0)$.

The reviewed above GL results for the fluctuation specific heat and conductivity are
only valid in the vicinity of the transition temperature for fields $H\ll
H_{c2}(0)$. In order to describe fluctuations close to the line $H_{c2}\left(
T\right)$ in its upper part, in the vicinity of the field $H_{c2}\left(  0\right)$
one has to develop the microscopic theory accounting for short-wave-length and
dynamical fluctuations
\cite{GL01,Lop0507,SSVG09,Fink09,GVV11a,TikhPRB12,TarasPRB13}. Both approaches for
the conductivity and heat capacity match in the wide domain of temperatures and
magnetic fields along the line $H_{c2}\left(  T\right) $ in its central part. In
contrast, for multiple-band superconductors, due to the possible diversity in the
partial coherence lengths, the GL description of fluctuations may break down at
magnetic fields \emph{significantly smaller} than $H_{c2}\left(  0\right)$ at
temperature quite close to $T_{c}$. In this case a proper microscopic description of
the short-wave-length fluctuation is necessary which we present in the next section.

\section{Fluctuation corrections in a two-band superconductor near the upper
critical field}

\begin{figure}[ptb]
\includegraphics[width=5in]{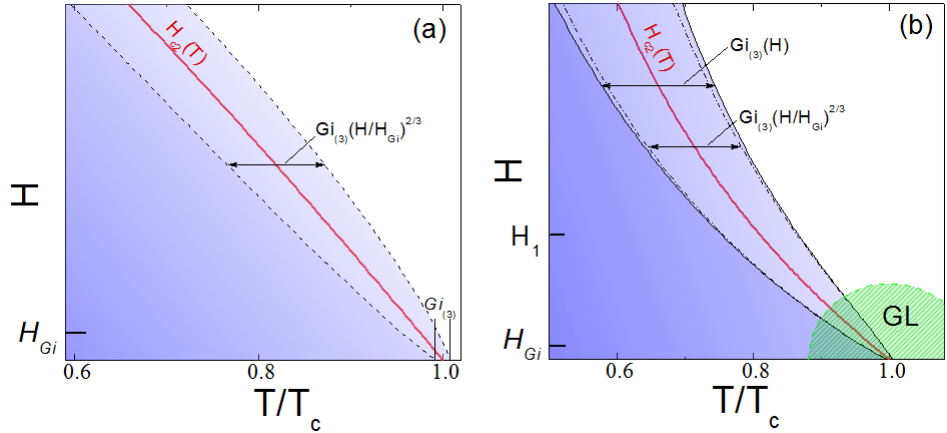}
\caption{Field dependences for fluctuation region for (a) single-band
and (b) two-band superconductors. $Gi_{(3)}(H)$ is the width of fluctuation
region given by (\ref{GiHTwoBand}) The shaded area near $T_c$ shows applicability
region of the conventional GL theory.}
\label{Fig-FluctWidth}
\end{figure}

\subsection{Model}

We consider a two-band superconductor described by the $2\times2$ coupling-constant
matrix $\lambda_{\alpha\beta}$. We assume strong scattering of quasiparticles inside
the bands (dirty limit) but neglect the interband scattering. In the case of very
different coherence lengths in different bands, the nonlocal effects become
important at moderate magnetic fields, significantly smaller than the
low-temperature critical field $H_{c2}(0)$. In this case the standard
Ginzburg-Landau model breaks down quite close to the transition temperature and more
general description is required. Fluctuation thermodynamics of a two-band
superconductor is described by the following quadratic nonlocal energy functional
for the band gap parameters $\Delta_{\alpha}$ \cite{GurVinPRL06}
\begin{eqnarray}
&  \mathcal{F}[\Delta_{\alpha}] =\min_{F_{\alpha}}\int d^{3}\mathbf{r}\left\{
\sum_{\alpha,\beta}\nu_{\alpha}\left(  w_{\alpha\beta}+\delta_{\alpha\beta
}\epsilon\right)  \Delta_{\alpha}^{\ast}\Delta_{\beta}\right.
\nonumber\\
&  +\left.  2\pi T\sum_{\alpha,\omega>0}\nu_{\alpha}\left[  \omega\left\vert
F_{\alpha}-\frac{\Delta_{\alpha}}{\omega}\right\vert ^{2}+\sum\nolimits_{j}
\frac{\mathcal{D}_{\alpha,j}}{2}|D_{j}F_{\alpha}|^{2}\right]  \right\}.
\label{EnerFunct}
\end{eqnarray}
Here $F_{\alpha}$ are the anomalous Green's functions, $\omega=\pi T(2m+1)$ are the
Matsubara frequencies, $\alpha=1,2$ is the band index, $j$ is the coordinate index,
$\mathcal{D}_{\alpha,j}$ are the diffusion coefficients (we assume that
$\mathcal{D}_{\alpha,x}=\mathcal{D}_{\alpha,y}$), and $D_{j}\equiv\nabla_{j}-(2\pi
i/\Phi_{0})A_{j}$. The degenerate matrix $w_{\alpha\beta}$ is defined as
\cite{KVV05}
\numparts
\begin{eqnarray}
w_{11}  &  \!=\!\frac{-\lambda_{-}/2\!+\!\sqrt{\lambda_{-}^{2}/4\!+\!\lambda
_{12}\lambda_{21}}}{\lambda_{11}\lambda_{22}-\lambda_{12}\lambda_{21}}
,w_{12}\!=\!-\frac{\lambda_{12}}{\lambda_{11}\lambda_{22}\!-\!\lambda
_{12}\lambda_{21}}\\
w_{22}  &  \!=\!\frac{\lambda_{-}/2\!+\!\sqrt{\lambda_{-}^{2}/4\!+\!\lambda
_{12}\lambda_{21}}}{\lambda_{11}\lambda_{22}-\lambda_{12}\lambda_{21}}
,w_{21}\!=\!-\frac{\lambda_{21}}{\lambda_{11}\lambda_{22}\!-\!\lambda
_{12}\lambda_{21}}
\label{wab}
\end{eqnarray}
\endnumparts
with $\lambda_{-}=\lambda_{11}-\lambda_{22}$. The functional (\ref{EnerFunct})
%is somewhat more general than the conventional Ginzburg-Landau
%one valid near the transition temperature $T_{c}$.
takes into account possible nonlocality when the spatial scale of the order
parameter variations becomes shorter than the one of the microscopic band coherence
lengths.

The anomalous Green's functions $F_{\alpha}$ rigidly follow fluctuations of the gap
parameters. Variation of the functional (\ref{EnerFunct}) with respect to
$F_{\alpha}^{\ast}$ gives the linear Usadel equations
\begin{equation}
\omega F_{\alpha}\!-\!\sum\nolimits_{j}\frac{\mathcal{D}_{\alpha,j}}{2}
D_{j}^{2}F_{\alpha}=\Delta_{\alpha}. \label{Usadel}
\end{equation}
Substitution of solution of this equation into (\ref{EnerFunct}) gives the
nonlocal functional for the gap parameters $\Delta_{\alpha}$.

To explore fluctuations in magnetic field, we will assume the Landau gauge,
$\mathbf{A}=(0,Hx,0)$ and introduce the eigenstates of the operator
$\sum\nolimits_{j}\mathcal{D}_{\alpha,j}D_{j}^{2}$ (Landau levels) as
\begin{equation}
\left[  -\frac{d^{2}}{dx^{2}}+\left(  q_{y}-\frac{2\pi}{\Phi_{0}}Hx\right)
^{2}+q_{z}^{2}/\gamma_{\alpha}^{2}\right]  \Psi_{n,\mathbf{q}}=a_{\alpha
,n,q_{z}}\Psi_{n,\mathbf{q}}
\end{equation}
with the eigenstates
\begin{equation}
 a_{\alpha,n,q_{z}}=\frac{2\pi H}{\Phi_{0}}(2n+1)+q_{z}^{2}/\gamma_{\alpha
}^{2}
\end{equation}
and the band anisotropy parameters
$\gamma_{\alpha}^{2}=\mathcal{D}_{\alpha,x}/\mathcal{D}_{\alpha,z}$. Expanding
$F_{\alpha}$ and $\Delta_{\alpha}$ with respect to these eigenstates,
$F_{\alpha}(\mathbf{r})=\sum_{n}\int\frac{d^{2}\mathbf{q} }{\left(  2\pi\right)
^{2}}F_{\alpha,n,\mathbf{q}}\Psi_{n,\mathbf{q}}$,
$\Delta_{\alpha}(\mathbf{r})=\sum_{n}\int\frac{d^{2}\mathbf{q}}{\left( 2\pi\right)
^{2}}\Delta_{\alpha,n,\mathbf{q}}\Psi_{n,\mathbf{q}}$, we immediately obtain from
(\ref{Usadel}) $F_{\alpha,n,\mathbf{q}} =\frac{\Delta_{\alpha,n,\mathbf{q}}}{\omega
+ \left(  \mathcal{D}_{\alpha ,x}/2\right)  a_{\alpha,n,q_{z}}}$. Substituting these
expansions into the functional (\ref{EnerFunct}), we arrive to the Gaussian energy
functional presented in terms of the fluctuating Landau-level amplitudes of the
order parameter
\begin{eqnarray}
\mathcal{F}  &  =\frac{1}{L_{x}}\sum_{n}\int\limits_{0}^{2\pi HL_{x}/\Phi_{0}
}\frac{dq_{y}}{2\pi}\int\limits_{-\infty}^{\infty}\frac{dq_{z}}{2\pi
}\nonumber\\
&  \nu_{\alpha}\left[  w_{\alpha\beta}+\delta_{\alpha\beta}\left(
\epsilon+\beta_{\alpha,n,q_{z}}\right)  \right]  \Delta_{\alpha,n,\mathbf{q}
}^{\ast}\Delta_{\beta,n,\mathbf{q}}. \label{NonLocGLFunct}
\end{eqnarray}
where
\begin{eqnarray*}
\beta_{\alpha,n,q_{z}}& \!=\!\beta\left[  \frac{2\pi\xi_{\alpha,x}^{2}H}
{\Phi_{0}}(2n\!+\!1)\!+\!\xi_{\alpha,z}^{2}q_{z}^{2}\right],\
\xi_{\alpha,i}^{2}(T)\!=\!\frac{\pi\mathcal{D}_{\alpha,i}}{8T} ,\\
\beta(x)  &  \equiv\psi\left(  1/2+2x/\pi^{2}\right)  -\psi\left(  1/2\right)
\label{beta1}\\
&  \approx\left\{
\begin{tabular}
[c]{l}
$x,\ \ x\ll1$\\
$\ln(8\gamma_{E}x/\pi^{2}),\ x\gg1$
\end{tabular}
\right. ,
\end{eqnarray*}
$\gamma_{E}\approx1.78$ is the Euler constant, and $\psi(x)$ is the digamma function.

Equation for the upper critical field \cite{Hc2Ref}, $H_{c2}$, is determined
by the condition of degeneracy of the matrix $w_{\alpha\beta}+\delta
_{\alpha\beta}\left(  \epsilon+\beta_{\alpha,0}\right)  $ with $\beta
_{\alpha,0}\!\equiv\beta_{\alpha,0,0}\!=\beta\left[  2\pi\xi_{\alpha,x}
^{2}H/\Phi_{0}\right]  $ giving
\begin{equation}
\left(  \epsilon\!+\!\beta_{2,0}\right)  w_{11}\!+\!\left(  \epsilon
\!+\!\beta_{1,0}\right)  w_{22}\!+\!\left(  \epsilon\!+\!\beta_{1,0}\right)
\left(  \epsilon\!+\!\beta_{2,0}\right)  \!=\!0. \label{Hc2Eq}
\end{equation}
The shape of the dependence $H_{c2}(T)$ following from this equation depends on the
relation between the bands coherence lengths $\xi_{\alpha,x}$ and relative strength
of superconductivity described by the matrix $w_{\alpha \beta}$.

\subsection{Single-mode approximation}

In zero magnetic field the superconducting instability develops in one channel
corresponding to the fixed relation between the band gap parameters,
$\Delta_{\alpha}\propto\psi_{\alpha}^{(1)}$, where $\psi_{\alpha}^{(1)}$ is the
eigenvector corresponding to zero eigenvalue of the matrix $w_{\alpha \beta}$,
$w^{(1)}=0$,
\[
w_{\alpha\beta}\psi_{\beta}^{(1)}=0
\]
or, explicitly%
\begin{eqnarray*}
\psi_{1}^{(1)} &  =\frac{w_{12}}{\sqrt{w_{11}^{2}+w_{12}^{2}}}=\mathrm{sign}(w_{12})\sqrt{\frac
{\nu_{2}w_{22}}{\nu_{1}w_{11}+\nu_{2}w_{22}}},\\
\psi_{2}^{(1)} &  =-\frac{w_{11}}{\sqrt{w_{11}^{2}+w_{12}^{2}}}=-\sqrt
{\frac{\nu_{1}w_{11}}{\nu_{1}w_{11}+\nu_{2}w_{22}}}.
\end{eqnarray*}
Here we used the relations $\nu_{1}w_{12}=\nu_{2}w_{21}$ and
$w_{11}w_{22}=w_{12}w_{21}$. In typical situation the second eigenvalue of this
matrix, $w^{(2)}=w_{11}+w_{22}$, is large meaning that this mode is strongly gapped,
and fluctuations in this channel can be neglected. That is why in this case one can
only take into account fluctuations in the unstable channel and use the projections
of the order parameter to the eigenstate of this channel $\psi _{\alpha}^{(1)}$. As
our purpose is to illustrate the multiband effects on fluctuations in the simplest
situation, we will limit ourselves with the single-mode approximation. In the
following we skip index $(1)$ in $\psi_{\alpha}^{(1)}$,
$\psi_{\alpha}^{(1)}\rightarrow\psi_{\alpha}$.

Equation for $H_{c2}$ becomes $\sum_{\alpha=1,2}\nu_{\alpha}\left(
\epsilon+\beta_{\alpha ,0}\right) \psi_{\alpha}^{2}=0$. One can show that for
arbitrary band-space vector $A_{\alpha}$ the identity
\[
\sum_{\alpha=1,2}\nu_{\alpha}A_{\alpha}\psi_{\alpha}^{2}=\nu_{1}\nu_{2}
\frac{A_{1}w_{22}+A_{2}w_{11}}{\nu_{1}w_{11}+\nu_{2}w_{22}}
\]
is valid. Being applied to the equation for $H_{c2}$, this identity allows to
transform it to the following simple form
\begin{equation}
\epsilon+\bar{\beta}_{0}=0, \label{Hc2SingleMode}
\end{equation}
where we introduced the band average $\bar{A}$ for the arbitrary $A_{\alpha}$ as
\begin{equation}
\bar{A}\equiv\frac{A_{1}w_{22}+A_{2}w_{11}}{w_{11}+w_{22}}. \label{AverDef}
\end{equation}
The weights with which the bands contribute to the averages are determined by the
strength of the superconducting pairing in them. For example, for stronger second
band $\lambda_{22}\gg \lambda_{11}$ we have $w_{11}\gg w_{22}$. Differentiating
(\ref{Hc2SingleMode}) with respect to the temperature, we obtain the useful relation
\begin{equation}
\overline{\beta_{0}^{\prime}\xi_{x}^{2}}\frac{2\pi}{\Phi_{0}}\left(
H_{c2}+TH_{c2}^{\prime}\right) =1. \label{dHc2dT}
\end{equation}
with $H_{c2}^{\prime}\equiv\left\vert dH_{c2}/dT\right\vert $. The temperature
dependence of the upper critical field becomes unconventional when the bands have
very different coherence lengths $\xi_{x}$, $\xi_{1,x}\gg\xi_{2,x}$. In this case in
the range $H_{1}=\frac{\Phi_{0}}{2\pi\xi_{1,x}^{2}}\ll H_{c2}\ll
H_{2}=\frac{\Phi_{0}}{2\pi\xi_{2,x}^{2}}$, equation for $H_{c2}$ takes the form
\[
\epsilon+\frac{\ln\left(  \frac{8\gamma_{E}H_{c2}}{\pi^{2}H_{1}}\right)
w_{22}+\frac{H_{c2}}{H_{2}}w_{11}}{w_{11}+w_{22}}=0.
\]
The characteristic two-band feature is a noticeable upward curvature of the
temperature dependence of the upper critical field at $H_{c2}(T)\sim H_1$, see Fig.\
\ref{Fig-FluctWidth}(b).

Taking the projection $\Delta_{\alpha,n,\mathbf{q}}=\Delta_{n,\mathbf{q}}
\psi_{\alpha}$, the nonlocal GL functional (\ref{NonLocGLFunct}) can be
transformed to the following form
\begin{equation}
\mathcal{F}=\frac{\nu_{\mathrm{av}}}{L_{x}}\sum_{n}\int\limits_{0}^{2\pi
HL_{x}/\Phi_{0}}\frac{dq_{y}}{2\pi}\int\frac{dq_{z}}{2\pi}\left(
\epsilon+\bar{\beta}_{n}\right)  |\Delta_{n,\mathbf{q}}|^{2}
\label{FGL-Nonloc}
\end{equation}
with
\[
\nu_{\mathrm{av}}=\frac{\nu_{1}\nu_{2}\left(  w_{11}+w_{22}\right)  }{\nu
_{1}w_{11}+\nu_{2}w_{22}}.
\]

\subsection{Specific heat}

The fluctuation correction to the free energy following from (\ref{FGL-Nonloc}) is given by
\begin{equation}
\delta F=-T\frac{H}{\Phi_{0}}\sum_{n}\int\frac{dq_{z}}{2\pi}\ln\frac
{A}{\epsilon+\bar{\beta}_{n,q_{z}}}. \label{FluctFreeEn}
\end{equation}
This gives the fluctuating specific heat $\delta C=-T\partial^{2}\delta
F/\partial T^{2}$
\begin{equation}
\delta C=\frac{H}{\Phi_{0}}\sum_{n}\int\frac{dq_{z}}{2\pi}\frac{\left(
1+T\partial\bar{\beta}_{n,q_{z}}/\partial T\right)  ^{2}}{\left(
\epsilon+\bar{\beta}_{n,q_{z}}\right)  ^{2}}. \label{SpecHeatGen}
\end{equation}
The temperature derivative in the numerator can be evaluated as $T\partial
\bar{\beta}_{n,q_{z}}/\partial T=-\frac{2\pi H}{\Phi_{0}}\overline
{\beta_{n,q_{z}}^{\prime}\xi_{x}^{2}}$. As in the case of a single-band
superconductor, the dominating divergency in the specific heat for $H\rightarrow
H_{c2}$ is given by the $n=0$ term. Using the expansion
\[
\epsilon+\bar{\beta}_{0}\approx2\pi\overline{\beta_{0}^{\prime}\xi_{x}^{2}
}\left(  H-H_{c2}\right)  /\Phi_{0}+\overline{\beta_{0}^{\prime}\xi_{z}^{2}
}q_{z}^{2},
\]
and relation (\ref{dHc2dT}), we find
\numparts
\begin{eqnarray}
\delta C  &  \approx\frac{H}{\Phi_{0}}\frac{\left(  TH_{c2}^{\prime}\right)
^{2}}{4\sqrt{\overline{\beta_{0}^{\prime}\xi_{z}^{2}}}\sqrt{H_{c2}
+TH_{c2}^{\prime}}\left(  H-H_{c2}\right)  ^{3/2}}\label{SpecHeatTwoBandSngH}\\
&  \approx\!\frac{H/\Phi_{0}}{4\sqrt{\overline{\beta_{0}^{\prime}\xi_{z}^{2}}
}\sqrt{1\!+\!H_{c2}/(TH_{c2}^{\prime})}\left[  \left(  T\!-\!T_{c2}\right)
/T\right]  ^{3/2}},\label{SpecHeatTwoBandSngT}
\end{eqnarray}
\endnumparts
where $T_{c2}\equiv T_{c2}(H)$ is defined by the relation $H_{c2}(T_{c2})=H$.
Multiband effects manifest themselves in this result via nonlinear temperature
dependence of the upper critical field and renormalization of the c-axis coherence
length $\xi_{z}\rightarrow\xi_{z,\mathrm{eff}}(H)=\sqrt{\overline{\beta_{0}^{\prime
}\xi_{z}^{2}}}$. In particular, in the range $H_{1}=\frac{\Phi_{0}}{2\pi\xi
_{1,x}^{2}}\ll H\ll H_{2}=\frac{\Phi_{0}}{2\pi\xi_{2,x}^{2}}$, we can explicitly
write
\[
\xi_{z,\mathrm{eff}}^{2}(H)=\frac{\left(  H_{1}/H\right)  w_{22}\xi_{1,z}
^{2}+w_{11}\xi_{2,z}^{2}}{w_{11}+w_{22}}.
\]
We can see that the contribution to  $\xi_{z,\mathrm{eff}}^{2}(H)$ from the large
coherence length decays in this region as $1/H$.

To evaluate the width of fluctuation region we have to compare the fluctuating
specific heat with the mean-field heat capacity jump. In the two-band case it is
given by $\Delta C=8\pi^{2}r_{12}\nu T_{c}/\left(  7\zeta(3)\right)  $, with the
renormalization factor\cite{SpecHeatJump}
\[
r_{12}=\frac{\nu_{1}\nu_{2}\left(  w_{11}+w_{22}\right)  ^{2}}
{\nu\left(  \nu_{1}w_{11}^{2}+\nu_{2}w_{22}^{2}\right)}\lesssim 1.
\]
Finally, we obtain
\begin{equation}
\frac{\delta C}{\Delta C}\approx\sqrt{Gi_{(3)}}\frac{H}{\tilde{H}_{c2}
(0)}\frac{\xi_{z}}{\xi_{z,\mathrm{eff}}}\frac{1}{\sqrt{1\!+\!H_{c2}
/(TH_{c2}^{\prime})}\left[  \left(  T\!-\!T_{c2}\right)  /T\right]  ^{3/2}}.
\label{DCTwoBandRel}
\end{equation}
where $\xi_{z}$ is the Ginzburg-Landau coherence length defined by
\[
\xi_{z}^{2}=\frac{w_{22}\xi_{1,z}^{2}+w_{11}\xi_{2,z}^{2}}{w_{22}+w_{11}}.
\]
Note that we absorbed the factor $r_{12}$ in the microscopic definition of
$Gi_{(3)}$ while the definition of $Gi_{(3)}$ in terms of the phenomenological
parameters remains unchanged.

Equation (\ref{DCTwoBandRel}) allows us to evaluate the field dependent width of the
fluctuation region defined by the criterion $\delta C(T,H)/\Delta C \sim 1$
\begin{equation}
Gi_{(3)}(H)\approx Gi_{(3)}^{1/3}\left(  \frac{H}{\tilde{H}_{c2}(0)}\right)
^{2/3}\left(  \frac{\xi_{z}}{\xi_{z,\mathrm{eff}}(H)}\right)  ^{2/3}.
\label{GiHTwoBand}
\end{equation}
One can see that, in-addition to the single-band broadening $Gi_{(3)}(H)\propto
H^{2/3}$ , a further smearing of the fluctuation region is caused by the reduction
of the $z$-axis coherence length. The fluctuation region  for a two-band
superconductor is illustrated in figure \ref{Fig-FluctWidth}(b).

\subsection{Conductivity}

The calculation of the fluctuation corrections to conductivity at arbitrary
temperatures and fields is a highly nontrivial problem due to pronounced nonlocal and
delay effects \cite{GL01,Lop0507,Fink09,GVV11a,TikhPRB12,TarasPRB13}. In contrast to
the Ginzburg-Landau region near the transition temperature, consensus on accurate
results for all contribution to conductivity is not achieved yet. In most part
of the phase diagram (except very low temperatures) dominating fluctuation
contribution to conductivity near the upper critical field is given by the
Aslamazov-Larkin term which was computed in the whole temperature range
\cite{GVV11a}. This result can be straightforwardly generalized to multiband case as
\begin{eqnarray}
&  \delta\sigma_{xx}^{\mathrm{AL}}(T,H)=\frac{e^{2}}{(2\pi)^{2}T}\sum
_{\alpha,\beta}\nu_{\alpha}\nu_{\beta}\sum_{n=0}^{\infty}(n+1)\int_{-\infty
}^{\infty}\frac{dq_{z}}{2\pi}\int_{-\infty}^{\infty}\frac{dz}{\sinh^{2}\left(
z/2T\right)  }\nonumber\\
\times &  \left\{  2\mathrm{Re}\left[  \Psi_{\alpha,n,n+1}\Psi_{\beta
n,n+1}\right]  \mathrm{Im}L_{\alpha\beta,n}^{R}\mathrm{Im}L_{\alpha\beta
,n+1}^{R}\right. \nonumber\\
+  &  \left.  \mathrm{Im}\left[  \Psi_{\alpha,n,n+1}\Psi_{\beta n,n+1}\right]
\left[  \mathrm{Im}L_{\alpha\beta,n}^{R}\mathrm{Re}L_{\alpha\beta,n+1}
^{R}+\mathrm{Im}L_{\alpha\beta,n+1}^{R}\mathrm{Re}L_{\alpha\beta,n}
^{R}\right]  \right\}  , \label{ALCondTwoBand}
\end{eqnarray}
where
\begin{eqnarray}
&  \Psi_{\alpha,nm}\equiv\beta_{\alpha,n,q_{z}}(-iz)-\beta_{\alpha,m,q_{z}
}(-iz),\\
\left(  L_{\alpha\beta,n}^{R}\right)  ^{-1}  &  =-\nu_{\alpha}\left[
w_{\alpha\beta}+\left[  \epsilon+\beta_{\alpha,n,q_{z}}(-iz)\right]
\delta_{\alpha\beta}\right],
\end{eqnarray}
and
\[
\beta_{\alpha,n,q_{z}}(z)\!\equiv\!\psi\!\left(  \frac{1}{2}\!+\!\frac
{z}{4\pi T}\!+\!\frac{4\xi_{\alpha,x}^{2}H(2n\!+\!1)}{\pi\Phi_{0}}
\!+\!\frac{2\xi_{\alpha,z}^{2}q_{z}^{2}}{\pi^2}\right)  \!-\!\psi\!\left(  \frac{1}
{2}\right)  .
\]
In the single-mode approximation we have
\begin{eqnarray}
&  L_{\alpha\beta,n}^{R}(-iz)=\psi_{\alpha}\psi_{\beta}L_{n}^{R}
(-iz),\ L_{n}^{R}=-\frac{1}{\nu_{\mathrm{av}}(\epsilon+\bar{\beta}_{n,q_{z}}
)},\\
&  \sum_{\alpha}\nu_{\alpha}\Psi_{\alpha,n,n+1}(-iz)\psi_{\alpha}^{2}
=\nu_{\mathrm{av}}\bar{\Psi}_{n,n+1}.
\end{eqnarray}
%we obtain
%\begin{eqnarray}
%&  \delta\sigma_{xx}^{\mathrm{AL}}(T,H)=\frac{e^{2}}{(2\pi)^{2}T}\sum
%_{n=0}^{\infty}(n+1)\int\limits_{-\infty}^{\infty}\frac{dq_{z}}{2\pi}
%\int\limits_{-\infty}^{\infty}\frac{dz}{\sinh^{2}\left(  z/2T\right)
%}\nonumber\\
%&  \times\left\{  2\mathrm{Re}\left[  \left(  \bar{\beta}_{n+1,q_{z}
%}(-iz)-\!\bar{\beta}_{n,q_{z}}(-iz)\right)  ^{2}\right]  \mathrm{Im}
%\frac{1}{\epsilon\!+\!\bar{\beta}_{n,q_{z}}(-iz)}\mathrm{Im}\frac
%{1}{\epsilon\!+\!\bar{\beta}_{n+1,q_{z}}(-iz)}\right.  \\
%+ &  \left.  \mathrm{Im}\left[  \left(  \bar{\beta}_{n+1,q_{z}
%}(-iz)\!-\!\bar{\beta}_{n,q_{z}}(-iz)\right)  ^{2}\right]  \left[
%\mathrm{Im}\frac{1}{\epsilon\!+\!\bar{\beta}_{n,q_{z}}(-iz)}
%\mathrm{Re}\frac{1}{\epsilon\!+\!\bar{\beta}_{n+1,q_{z}}(-iz)}
%\!+\!\mathrm{Im}\frac{1}{\epsilon\!+\!\bar{\beta}_{n+1,q_{z}}
%(-iz)}\mathrm{Re}\frac{1}{\epsilon\!+\!\bar{\beta}_{n,q_{z}}
%(-iz)}\right]  \right\}  .\nonumber
%\end{eqnarray}
In the classical regime we can use the small-frequency expansions, $\sinh\left(
z/2T\right)  \approx z/2T$, $\bar{\beta}_{n,q_{z}}(iz)=i\eta_{n}z+\bar
{\beta}_{n,q_{z}}$ with $\bar{\eta}_{n}=\frac{\pi}{8T}\bar{\beta} _{n,0}^{\prime}$.
Performing the frequency integration in (\ref{ALCondTwoBand}) one finds
\[
\fl
\delta\sigma_{xx}^{\mathrm{AL}}\!=\!\frac{2}{\pi}Te^{2}\!\sum_{n=0}^{\infty
}(n\!+\!1)\!\int\limits_{-\infty}^{\infty}\!\frac{dq_{z}}{2\pi}\frac{\bar{\eta}_{n}
\bar{\eta}_{n+1}\left(  \bar{\beta}_{n,q_{z}}\!-\!\bar{\beta}_{n+1,q_{z}
}\right)  \left[  \left(  \epsilon\!+\!\bar{\beta}_{n,q_{z}}\right)\!  /\bar{\eta
}_{n}^{2}\!-\!\left(  \epsilon\! +\! \bar{\beta}_{n+1,q_{z}+1}\right)\!  /\bar{\eta}
_{n+1}^{2}\right]  }{\left(  \epsilon\!+\!\bar{\beta}_{n,q_{z}}\right)  \left(
\epsilon\!+\!\bar{\beta}_{n+1,q_{z}}\right)  \left[  \left(  \epsilon\!+\!\bar{\beta
}_{n,q_{z}}\right)  /\bar{\eta}_{n}\!+\!\left(  \epsilon\!+\!\bar{\beta
}_{n+1,q_{z}}\right)  /\bar{\eta}_{n+1}\right]  }.\nonumber
\]
In the vicinity of the $H_{c2}(T)$ line, as usual, we can keep only the $n\!=\!0$
term and approximately obtain
\begin{equation}
\delta\sigma_{xx}^{\mathrm{AL}}  \approx\frac{e^{2}}{4}\int_{-\infty
}^{\infty}\frac{dq_{z}}{2\pi}\frac{\bar{\beta}_{0}^{\prime}}{\epsilon
+\bar{\beta}_{0,q_{z}}}\nonumber.
\end{equation}
Expanding $\bar{\beta}_{0,q_{z}}\approx\bar{\beta}_{0}+\overline{\beta
_{0}^{\prime}\xi_{z}^{2}}q_{z}^{2}$ and performing $q_{z}$ integration, we
finally obtain
\numparts
\begin{eqnarray}
\delta\sigma_{xx}^{\mathrm{AL}}  &  \approx\frac{e^{2}
\overline{\beta_{0}^{\prime}}}{8}\sqrt{\frac{H_{c2}+TH_{c2}^{\prime}}{\left(  H-H_{c2}
\right)  \overline{\beta_{0}^{\prime}\xi_{z}^{2}}}}\label{SigxxDominH}\\
&  =\frac{e^{2}\overline{\beta_{0}^{\prime}}}{8}\sqrt{\frac{T\left[
1+H_{c2}/\left(  TH_{c2}^{\prime}\right)  \right]  }{\left(  T-T_{c2}\right)
\overline{\beta_{0}^{\prime}\xi_{z}^{2}}}}.
\label{SigxxDominT}
\end{eqnarray}
\endnumparts
We can see that, similar to the specific-heat correction
(\ref{SpecHeatTwoBandSngH},\ref{SpecHeatTwoBandSngT}), the multiband effects
influence these results  via  nonlinear temperature dependence of $H_{c2}$ and
renormalization of the c-axis coherence length. An additional factor
$\overline{\beta_{0}^{\prime}}$ appears due to renormalization of the dynamics
coefficient $\eta_{0}$.

To evaluate the apparent width of the transport transition, we normalize
$\delta\sigma_{xx}^{\mathrm{AL}}$ to the normal-state conductivity $\sigma
_{n}=2e^{2}(\nu_{1}\mathcal{D}_{1}+\nu_{2}\mathcal{D}_{2})$ and obtain%
\begin{equation}
\frac{\delta\sigma_{xx}^{\mathrm{AL}}}{\sigma_{n}}\approx\frac{\pi^{4}%
\sqrt{Gi_{(3)}}\overline{\beta_{0}^{\prime}}}{14\zeta(3)}\frac{r_{12}\nu \xi
_{x}^{2}}{\nu_{1}\xi_{1,x}^{2}+\nu_{2}\xi_{2,x}^{2}}\frac{\xi_{z}}%
{\xi_{z,\mathrm{eff}}}\sqrt{\frac{T}{T-T_{c2}}}.
\end{equation}
We can see that, in comparison to the similar single-band result
(\ref{sigRatSnglBnd}), this ratio contains several additional factors,
$\overline{\beta_{0}^{\prime}}$, $\xi _{z}/\xi_{z,\mathrm{eff}}$, $r_{12}$, and
$\xi_{x}^{2}\nu/\left(  \nu_{1} \xi_{1,x}^{2}+\nu_{2}\xi_{2,x}^{2}\right) $. All
these factors are smaller than one. The most important among them is the last
factor. The Ginzburg-Landau coherence length in its numerator $\xi_{x}$ is
determined by the bands contributions weighted by the relative strength of
superconductivity, $\xi_{x}^{2}=\left(
w_{22}\xi_{1,x}^{2}+w_{11}\xi_{2,x}^{2}\right) /\left( w_{22}+w_{11}\right)$, while
the normal-state average in the denominator $\left(
\nu_{1}\xi_{1,x}^{2}+\nu_{2}\xi_{2,x}^{2}\right)/\nu$ is just determined by the
partial densities of states. This ratio may be very small if the band with large
coherence length has weak superconductivity. In other words, the relative
fluctuation correction reduces because the band with strong superconductivity may
dominate the fluctuation correction, while the band with largest mobility dominates
the normal-state conductivity. In this case the apparent width of the transport
transition narrows down and shifts to lower temperatures in comparison with the
thermodynamic transition. Even though this effect exists already in a single-band
case at sufficiently high fields, as discussed in Section \ref{Sec-SigSnglBand}, we
see that it may be strongly enhanced in the multiband case.

\section{Conclusions}

We considered the fluctuating specific heat and conductivity in the vicinity of the
upper critical field line for a two-band superconductor in dirty limit.
Multiple-band effects strongly influence superconducting properties when the bands
have very different coherence lengths. In the case of strongly different bands the
transition to superconductive state is mainly determined by the properties of the
rigid condensate of the ``strong'' band, while the large coherence length of the
Cooper pairs from the ``weak'' band leads to the nonlocality in fluctuation
behaviour and break down of the simple GL picture. This large coherence length
$\xi_1$ sets the field scale $H_1$ at which the upper critical field has upward
curvature. The contribution of the ``weak'' band to the fluctuation corrections
rapidly decreases above $H_1$. As expected, the multiple-band effects do not change
the functional forms of dominating divergencies of the fluctuating corrections for
$H\rightarrow H_{c2}(T)$, $\delta \sigma \propto (H-H_{c2})^{-1/2}$ and $\delta C
\propto (H-H_{c2})^{-3/2}$. The corresponding coefficients are modified by the local
slope of the upper critical field, $dH_{c2}/dT$, and renormalization of the z-axis
coherence length.

Analyzing known results for fluctuating specific heat and conductivity in a
single-band superconductor, we observed that for  strong enough fields $H\gg
194H_{Gi}$ the relative fluctuation correction to conductivity remains small even
when fluctuations of the order parameter become strong and non-Gaussian.
Qualitatively, this theoretical finding corresponds to the experimental observation
that in some iron-based superconductors the resistive transition at high magnetic
field takes place at apparently lower temperatures in comparison with the
specific-heat transition \cite{SerafinPRB10,WelpMarcenat}. For example, for the
compound BaFe$_{2}$(As$_{0.7}$P$_{0.3})_{2}$ using $\tilde{H}_{c2}(0)\approx 67.5$T
and $Gi_{(3)}\approx 3\times 10^{-5}$,
%\cite{pnictideGis},
we obtain that such behaviour is expected above the field $194H_{Gi}\sim 0.37$T.
This shift of the apparent transport transition to lower temperatures may be further
enhanced by the multiple-band effects, because the ``weak'' band may dominate the
normal-state conductivity, while its contribution to the fluctuation correction is
reduced due to weakness of superconductivity.

%In this paper we considered the dirty-limit case which is realized when scattering
%inside the bands is very strong. In this situation the microscopic theory has the
%simplest form. In fact, due to very small coherence lengths, most iron-based
%superconductors are actually in clean limit which is not quantitatively described by
%the presented theory. Nevertheless, in the case of large difference between the
%coherence lengths in different bands, a general picture for the clean case is
%similar. Namely, in this case strong nonlocality in the band with the large
%coherence length also will lead to breaking of the conventional behaviour of
%fluctuations above the field scale set by this length.

In this paper we considered the dirty-limit case which is realized when scattering
inside the bands is strong and the microscopic theory has the simplest form. In
fact, due to very small coherence lengths, most iron-based superconductors are
actually in clean limit which is not quantitatively described by the presented
theory. Nevertheless, we believe that this circumstance does not change our results
qualitatively. Indeed, as it was pointed above, along the line $H_{c2}(T)$, for not
very low temperatures the dominant contribution to fluctuation conductivity is the
Aslamazov-Larkin one. In the framework of the standard local theory it can be
obtained in the hydrodynamic limit and all information concerning the impurities
concentration is included in the coherence lengths (see equations (\ref{C3D}) and
(\ref{sxx})). Transition to the clean case results in decrease of fluctuation
effects inversely proportional to growing coherence lengths.
%
%Yet, it does not result in change of the established above hierarchy of the
%fluctuation contributions: vice versa, Maki-Thompson and DOS contributions gradually
%compensate each other ( D. V. Livanov, G. Savona, A. A. Varlamov, PRB 62
%8675(2000)), while the Aslamazov Larkin term remains of the same form (A.G.Aronov,
%S.Hikami, A.I.Larkin, PRB 51 3880 (1995)).
What concerns the two-band superconductors, in the case of large difference between
the coherence lengths in different bands, a general picture for the clean case
remains similar to the dirty one. Namely, in this case strong nonlocality in the
band with the large coherence length also will lead to breaking of the conventional
behaviour of fluctuations above the field scale set by this length.

\ack

The authors would like to thank U. Welp and C. Marcenat for discussion of
experimental data for P-doped 122 iron pnictide. Work by A.~E.~K. was supported by
the U.S. Department of Energy, Office of Science, Materials Sciences and Engineering
Division. A.~A.~V. acknowledges support of the European FP7 program, Grant No.\
SIMTECH 246937.

\end{document}